**The State of Applying Artificial Intelligence to Tissue Imaging for Cancer Research and Early Detection**

Michael Robben, PhD[1,2], Amir Hajighasemi[1,2], Mohammad Sadegh Nasr[1,2], Jai Prakesh Veerla[1,2], Anne Marie Alsup[1,3], Biraaj Rout[1,2], Helen H. Shang, MD[1,4], Kelli Fowlds[1,3], Parisa Boodaghi Malidarreh[1,2], Paul Koomey[1,2], MD Jillur Rahman Saurav[1,2], Jacob M. Luber, PhD[1,2,3,5]

[1]Department of Computer Science and Engineering, The University of Texas at Arlington

[2]Multi-Interprofessional Center for Health Informatics, University of Texas at Arlington

[3]Department of Bioengineering, The University of Texas at Arlington

[4]Department of Internal Medicine, Ronald Reagan University of California Los Angeles Medical Center

[5]Department of Biology, The University of Texas at Arlington

## Abstract

Artificial intelligence represents a new frontier in human medicine that could save more lives and reduce the costs, thereby increasing accessibility. As a consequence, the rate of advancement of AI in cancer medical imaging and more particularly tissue pathology has exploded, opening it to ethical and technical questions that could impede its adoption into existing systems. In order to chart the path of AI in its application to cancer tissue imaging, we review current work and identify how it can improve cancer pathology diagnostics and research. In this review, we identify 5 core tasks that models are developed for, including regression, classification, segmentation, generation, and compression tasks. We address the benefits and challenges that such methods face, and how they can be adapted for use in cancer prevention and treatment. The studies looked at in this paper represent the beginning of this field and future experiments will build on the foundations that we highlight.

## Introduction

Tissue section imaging has been used for over 100 years in the diagnosis and treatment of cancer and has evolved from simple stains to the use of Hematoxylin and Eosin (H&E) and Immunohistochemistry (IHC) to resolve tissue and cellular structures (Harrison and Woolner, 1963; Geschickter et al., 1934). Cancer can be both detected and staged through the use of morphological and peptide markers (King and King, 1986; Kerr et al., 2014). Currently, tissue pathology is used in conjunction with radiological images to allow for the rapid detection of solid tumors in non-invasive tests, however, prognostic and therapeutic determination rely on the biopsy and microscopic imaging of tumors (ASPE panel, 2010).

The microscopic imaging of a sectioned tumor biopsy can reveal important information including confirmation of malignancy, stage of invasion, and possible response to certain therapeutics (Kerr et al., 2014). Cancer staging and prognostic evaluation by pathological experts has been found to have as low as

67% consensus in some forms of Non lymphoblastic lymphomas (Freedman and Manchin, 1993). This has increased with newer techniques and computer assisted visualization but still relies on an unreliable subjective human determination of disease. Older methods of computer assisted diagnosis and image analysis have been well reviewed by Gurcan et al. (2009). While they have had a tangible effect on diagnostic success rates, they are not able to replicate the flexibility and accuracy seen in newer forms of computer vision.

Computer vision began as simple experiments in the 1960's to try to reproduce human visual recognition in computers and has quickly expanded into one of the fastest growing fields of Artificial Intelligence (AI) research today (Fukushima, 1980). In the period following the initial computer vision models, many labs attempted to apply computer vision algorithms in tissue image analysis in cancer (Erler et al., 1994; Stotzka et al., 1995; Tewari and Perinchery, 1998), but most suffered from low accuracy and generalizability. This was a pervasive problem with most computer vision algorithms of the time, leading to a general error rate of more than 26% for an entire decade (LeCun, 1998). However, computer vision really took off in 2012 with the publication of Alexnet (Krizhevsky et al., 2012) which reduced the error rate of image classification to 16.4%. This method improved on the design of previous models, creating a stacked deep convolutional network using the ReLu activation function and captured important features of the images. This has led to the development of even higher performing computer vision models like VGG (Simonyan et al., 2014) and Resnet (He et al., 2016), which have pushed the field to very large parameter models with error rates of less than 5%.

Due to this surge of high accuracy models, the cancer tissue imaging research community has begun applying computer vision algorithms to solve problems in image analysis and diagnosis (Sharma et al., 2017; Mishra et al., 2018). However, the modern application and design of computer vision algorithms in biomedical tissue analysis has not yet been exhaustively documented.

In our review of current methodologies used in analysis and diagnostics we find that AI algorithms can be categorized into 5 discrete "tasks" in terms of purpose and design (Fig 1a). Regressive tasks which seek to predict a value from image data, classification tasks which label an image or parts of an image, generative tasks that create new images or data from tissue images, segmentation tasks which demarcate an image based on underlying features, and compressive tasks which seek to reduce the size of an image for storage or indexing. In this review, we will characterize current methodologies and models for each of these tasks and discuss their current use in cancer tissue imaging.

## Data Collection and Feature Extraction

Training AI models on tasks in cancer imaging is arguably easier than ever before, with large repositories of data enabled by digitalization of hospital pathology data and electronic health records. However, model quality is still highly dependent upon how the tissue is prepared and imaged and the quality of features that are passed to the models. Preparation of tissue imaging data involves either frozen or paraffin embedded tumor samples sliced into 5-20 µM sections and imaged under a microscope after a contrastive stain is applied to visualize normally transparent cellular structures (Kerr et al., 2014). Image training data can range from single pictures of a part of a tissue mount to a whole slide image constructed from sequential image patches stitched together and can be greater than 10,000 pixels in height and width depending upon microscope magnification. This provides computational scientists with a wealth of training data, as tissue pathology is a commonly used tool in the diagnosis and evaluation of cancer progression, and patient imaging data is generally stored for many years by hospitals (Schüffler et al., 2021).

The most common tissue preparation is using an H&E stain which uses colored dyes to produce differently colored cell nuclei and cytoplasm. These are the most abundant cancer image types and will often represent important characteristics about the cancerous tissue, such as the general tissue morphology and cell shape. Morphological features that can be used by both pathologists and ML algorithms to classify cancerous tissue include nuclear position and density (Bhattacharjee et al., 2021 and Kwak et al., 2017), general tissue texture (e.g., rough vs smooth) (Kather et al., 2016 and Poojitha and Sharma, 2019), and tissue specific qualities like glandular uniformity (Nir et al., 2018).

One tool that pathologists use to stage and diagnose cancers is antibody staining of commonly found protein markers (Singh and Mukhopadhyay, 2011). IHC staining can be very useful to pathologists due to the differential expression of protein markers in different cancer types and stages (Prasad et al., 2005 and Atkins et al., 2004). This allows pathologists to make clinical decisions about patient care using more information than just morphology. This greatly affects the development of AI models. While a model that is developed for morphological features from an H&E slide can be applied to more samples, models trained on stained slides have been found to have a higher accuracy in classification and regression tasks (Table 1). This is likely because protein expression features share more relevance to cancer development than morphological features.

Extracting features from an H&E or IHC stain are necessary for many types of machine learning models, such as linear regressions or random forests. For many newer computer vision models, however, extracting features of large image files prior to training is unnecessary as deep learning networks that use convolutional layers can train on raw image input (Ronneberger et al., 2015). A convolutional network can take a pixel field as input and create a new vector representing underlying patterns by passing a sliding filter which sums or averages the RGB values (Fig. 1b). Extracting features from a WSI may be unnecessary and could add undue computational load to models that are trained on millions of WSI, as was demonstrated by Cheng et al., (2018), who used a CNN + Recurrent Neural Network (RNN) to classify cervical cancer at a rate of 88.5%. However, studies have shown that whole slide image (WSI) inputs are susceptible to low accuracy and poor generalization in different parts of tissues which can be remedied by training models on tissue patches that contain sub tissue structures relevant to the task (Hou et al., 2016). This, however, presents a problem for training data, as most slide label data is given for an entire slide rather than sections of a slide. Some authors have suggested adopting weakly supervised learning methods such as Multiple Instance Learning (MIL) which splits the slide image into many "bags" each of which can be trained on multiple labels to identify deeper label features (Campanella et al., 2019 and Lu et al., 2021).

Sometimes the use of alternative imaging modalities can result in the production of features important for ML task training. Advanced imaging techniques, such as IR spectral imaging of tissue sections can produce more important features in cancer tissue analysis than H&E or IHC, such as metabolic identifiers between common cancers (Diem et al., 2004). This was demonstrated by Krafft et al., (2006) when they successfully deployed a Soft Independent Model of Class Analogies (SIMCA) supervised classifier to predict primary cancer site in brain metastases which share too many morphological similarities between different adenocarcinomas. Medical imaging data may not even be necessary for training model weights as res-nets pre trained on large image databases are an intuitive way to partition features in a tissue image, as has been demonstrated previously (Alexandre et al., 2019).

## Segmentation

Histopathological analysis and diagnosis of disease occurs at the tissue and cellular level. Certain diseases can be identified through stains, while others require comparison of tissue and cell morphology to reference healthy tissue slides (Gurcan et al., 2009). Due to the variation in tissue preparation and imaging, subjective analysis by pathologists incurs high rates of error in evaluation of disease or the annotation of disease regions on slides (Raab et al., 2005 and Middelton et al., 2014). There is mounting interest from both the research and clinical sides in computer assisted diagnostic (CAD) programs as tissue and cell segmentation models have been developed to assist with the extraction of relevant clinical information from whole slide images.

Cell segmentation involves segmenting the image for each individual cell that makes up the overall tissue. Making observations on the cellular level as opposed to the tissue level can have important effects, indicated by one study that showed a per cell effect on survival for biomarker prediction (Cui et al., 2020). Unique applications using segmented cancer cells can be seen in models used to detect mitotic cells or cancer stem cells (Lu et al., 2021; Aichinger et al., 2017). The primary method for cell segmentation since the digitization of microscope images has been thresholding pixel value intensities to separate cells from the background and then graph based methods to identify individual nuclei (Ortiz de Solorzano et al., 1999). This method generally performs well on most tissue images that use either H&E or IHC, with newer methodologies able to match nearly 90% accuracy to hand segmented images (Al-Kofahi et al., 2009). These techniques rely on the ability to distinguish nuclear boundaries either through brightfield stains like Hematoxylin or fluorescent stains like DAPI, because nuclei border morphology stays relatively constant between cells as opposed to cell membrane morphology. This can be a problem, however, in cancer tissues, in which nuclear and membrane morphology varies greatly (Di Cataldo et al., 2010).

One of the primary challenges in cell and nuclear segmentation is image overlap of nuclei belonging to different cells. Some methods have been proposed to address this, such as one study which used CNNs to predict the boundary of the nuclei and then subtract that from the nuclear layer to obtain a size reduced non overlapping nuclear segmentation (Vu et al., 2019). While nuclear segmentation can reliably inform on cell number and density, segmentation of the shape of the cell provides important pathological information and current methodologies that apply Voronoi or watershed image division cannot accurately capture cell morphology (Greenwald et al., 2022). This becomes more important as new methods are developed to capture highly multiplexed protein expression in a spatial context (Tsujikawa et al., 2019).

The use of deep learning models in cell segmentation has been lock-step with computer vision algorithm development. U-Net, a type of Fully Convolutional Network (FCN), was created in 2015 for the explicit purpose of cell segmentation (Fig 2a) (Ronneberger et al., 2015). U-net won the ISBI cell tracking challenge in 2015, achieving accuracies of 97% and 77% on each of the unstained test datasets. It has been cited 10,000 times and serves as the basis for many advanced deep learning-based cell and biomedical image segmentation models. A recent benchmark study of deep CNN models for segmentation of cells from breast cancer tissue images identified that U-Net architectures perform best for this purpose (Lagree et al., 2021). The further development of stardist (Schmidt et al., 2018) and cellpose (Stringer et al., 2021) have provided researchers and pathologists with practical implementations of such models. Newer models continue to be developed such as those using res-nets with larger and differently annotated training data sets to segment both nuclei and cytoplasm (Greenwald et al., 2022). Some models have been designed to take advantage of mask R-CNNs which were developed to segment individual objects in regular computer vision in order to individually segment cells (Lee et al., 2022; Jung et al., 2019).

As cancer cells show greatly variable morphologies it is important to develop cell segmentation algorithms that can accurately accommodate offending tissues. One of the most recently published studies for nuclear segmentation of breast cancer cells using UNET++ architectures was only able to achieve a 0.80 F1 accuracy score (Dinh et al., 2021). This represents the average of most of these algorithms and is still far off from the reported 4.87-11.80% error rate in pathological examination of cancerous tissue by hand (Raab et al., 2005). The poor performance seen in these studies may possibly be due to key structural differences in cancer tissue, including abnormal membrane and nuclear morphology, increased cell density and a high mixture of cell composition in a 3d tumor environment (Litjens et al., 2016). One promising new technology for segmentation of cancerous cells is Visual Transformers (ViT) that are built off of attention heads (Fig 2b) which can understand translatable features between samples in a way FCNs and RNNs cannot (Barzekar et al., 2023; Vaswani et al., 2017). While there may be some differences for cancer cell segmentation that model interoperability will need to solve, many cell segmentation models trained on a mix of uncancerous and cancerous tissue are generalized enough to work with cancer tissues specifically.

Perhaps more important for clinicians is the ability to segment large regions to classify tumors and metastatic growths as well as other tissue components like tertiary lymph nodes and vascular invasion into a tissue. While cell morphology and distribution are quite important markers for pathologists, tissue segmentations would provide them with details more useful to the prognostic indicators of patient health (Huang et al., 2020). Tissue segmentation algorithms utilizing U-net, res-net, or some combination of the two have been developed to identify regions of melanoma (Thomas et al., 2021), mask and classify lymph node metastases (Khened et al., 2021), and predict angiogenic blood vessel growth into cancerous tissue (Yi et al., 2019). Tissue segmentation may eventually prove to be an integral part of the pathology workflows. Computer Assisted Diagnostic (CAD) models could help pathologists distinguish between different types of breast cancer, which require a completely different treatment regimen (Cruz-Roa et al., 2014). Segmented cancer tissue images are also an ideal starting point for integration with other classification and regression models such as Support Vector Machines (SVM) (Tabibu et al., 2019; Di Cataldo et al., 2008).

While their use in CAD and surgical recommendations is apparent, tissue segmentation models suffer from low clinical relevance due to low generalizability and inaccuracies compared to pathologist mediated segmentation. An accuracy of less than 90%, barely outscoring more basic thresholding algorithms, is not an uncommon occurrence in these segmentation models. Increases in model performance in the last 10 years has been driven by grand challenges in segmentation tasks offered by societies with a vested interest in medical cancer pathology, such as the International Conference on Pattern Recognition (ICPR) and Medical Image Computing and Computer Assisted Intervention (MICCAI) conferences (Foucart et a., 2022). U-Net, which was developed in 2015, still holds a top scoring position in the GLaS competition for glandular segmentation of colorectal adenocarcinoma tissue images. In the Data Science Bowl challenge data set, accuracy has not moved past 92% dice accuracy (Tang et al., 2022). Many segmentation models seem to overfit and not generalize well to new data. Ho et al., (2021) proposed a multiple magnification FCN architecture for cancer tissue segmentation, and while it creates more smooth segmentations it seems to unintentionally over smooth and segment regions that are not concurrent with ground truth. ViT models have managed to slightly improve model accuracy since their discovery in 2017 (Tang et al., 2022; Wazir and Fraz, 2022).

## Regression and classification

Using machine learning models in pathological analysis of cancer has mostly taken the form of classification or regression models that are used to predict a diagnosis or correctly stage a cancer for prognostic purposes. These models are typically trained in a supervised manner as data can readily be taken from patient imaging and Electronic Health Records (EHR). The proliferation of large data repositories such as the National Institute for Health Cancer Genome Atlas (NIH TCGA) allow for large models to be generated operating off of thousands of patients EHR, imaging data, and sequencing results (Shang et al., 2023). In fact, through large multimodal studies enabled by TCGA, we have learned that sequencing and imaging data can build on each other to achieve a nearly 0.99% accurate c-index prediction in thyroid, kidney and colon adenocarcinomas (Vale-Silva and Rohr, 2021).

Models that can regressively predict numerical values are important for learning information related to patient survival and treatment effectiveness such as c-index, tumor mass, growth, and invasion. Regressive models for cancer survival have been previously built using a variety of measurable patient factors including gene expression (Zhao et al., 2021), volatile chemical fingerprinting (Ryu et a., 2020), and clinical risk factors (Arem and Loftfield, 2018). More recent models have been developed to assess survivability based on medical imaging, primarily through MRI and PET scans. While radiological imaging alone is a decent indicator of survival or response to treatment, Shao et al., (2020) demonstrated that combining features from radiological and pathological imaging modalities result in >85% accuracy for predicting such values. It appears that regressive information about survival and prognosis correlate highly (~90% accuracy) with features that can be extracted from histopathology images, such as cell morphology (Cui et al., 2020) and tumor edge shape (Wang et al., 2018). This is possibly due to tissue level morphological and cell composition differences in late stage and aggressive tumors (Baghban et al., 2020). Studies have also shown that combining clinical risk factors with pathological imaging data increases accuracy of survival predictions through c-index metrics (Zhu et a., 2016). This was demonstrated more recently in a study performed on H&E data from The Cancer Imaging Archive (TCIA), where a patch sampled and pooled FCN trained to predict survival based on histopathology and risk factors performed better than the model predicting survival based on risk factors alone (Wulczyn et al., 2020).

Classification models are built to predict everything from whether or not a patient has cancer to what kind of cancer that patient has. Classification models are very useful in tissues where multiple cell types can develop into different but similar presenting cancers such as lung adenocarcinoma and squamous cell carcinoma which are treated differently and have different prognoses (Coudray et al., 2018). A model developed on lung H&E slides was able to classify the correct disease type with an AUC of 0.99. Both res-nets and U-nets have proven to be invaluable models for histopathological based tissue classification and regression, allowing researchers and clinicians to do very difficult things like identify the presence of tertiary Lymph Nodes, which are indicative of a poor prognosis (Wulczyn et al., 2021 and Harmon et al., 2020). As the field of AI research advances it is important to constantly include the latest algorithms in practical models to continually iterate on accuracy. Using dropout with a Bayesian Neural Network (BNN) improved the AUC from 0.02 to 0.99 in diagnosing follicular lymphoma vs hyperplasia

(Syrykh et al., 2020). However, latest doesn't always mean the best performance, as a cursory comparison of existing models with ViT found that well optimized res-nets like ConvNeXt beat high performing transformers like SwinT on classification accuracy in most cancer types (Springenberg et al., 2022). Cancer scoring, which is a relative marker of cancer invasiveness and a general yardstick for patient survival before treatment, is an interesting case in terms of predictive modeling. Although it is considered a classification problem, many features of grading are continuous but when predicted in a regressive model, results in accuracies lower than pathologist classification (Wetstein et al., 2022; Wang et al., 2022; and Nagpal et al., 2019).

## Generation

Generative models utilize deep learning networks to create new data that represents but is not identical to real world data. This can be used for tasks as innocuous as creating photographs of people who do not exist (Karras et al., 2019); however, more researchers are realizing its uses in biomedical imaging research. Researchers from many labs have shown that rather than human faces, Generative Adversarial Networks (GAN) can be used to produce clinically accurate data like pictures of malignant melanomas (Baur et al., 2018), mammograms of invasive ductal adenocarcinomas (Wu et al., 2018), and chest CTs of different types of lung cancer (Ryo et al., 2022). These models work off of different methods, but the earliest GANs take a generative network such as pix2pix (Isola et al., 2017), which can input noise and output a random image and trained them using a discriminative model which would generate an error loss to real images.

Wong et al., (2020) used such a model to generate highly realistic synthetic tissue images of breast adenocarcinoma, particularly benign samples, to better train classification models due to a low prevalence of existing sample datasets. There is still discourse around the efficacy of using generated images to augment pre-existing data in a medical context as there is no existing method to evaluate the reliability and reproducibility of data that doesn't exist. However, this use of the models can be very effective to better develop missing data and edge cases, such as very noisy or occluded data (Chlap et al., 2021). Generative models may also be able to add clinically relevant data to old samples. Saurav et al., (2022) demonstrated that conditional GANs guided by feature selection methods can be used to generate missing protein data from a multiplexed IHC, greatly expanding the diagnostic capabilities of previously or sparsely collected IHC data.

A clear use case for GAN in current pathology practices is super resolution, or models that can improve the resolution of medical imaging data. Super resolution generative models have been applied with great effect to many technologies including CT, MRI, and ultrasound data (Zhu et al., 2022; Ahmad et al., 2022; and Nneji et al., 2021). This kind of model has major implications in pathology and biomedical research using histological tumor slides, as images taken with low resolution are very common and may not be able to resolve subcellular or even at times cellular structures that are important to diagnosis and biological discovery (Shahidi, 2021).

Transfer learning is quickly becoming very important to analysis of pathological cancer imaging data. Several studies have shown that models pretrained on classical image libraries like ImageNet can be used to segment features and classify tumors at greater than 80% accuracy (Bungărdean et al., 2021 and Kim et al., 2020). Other applications purport the ability to predict molecular markers of cancer from simple

H&E-stained slides which would allow pathologists to obtain more clinical information from a single sample (Su et al., 2022).

## Compression

The universal practice of biopsy preservation by formalin fixation and paraffin embedding (FFPE) has led to the accumulation of tens of thousands of cancer tissue blocks across medical institutions. Across the country this can account for petabytes of histological data for use in cancer image analysis and model development (Yu and Zhu, 2010). For example, the tissue bank at the MD Anderson Medical Center held 25,000 FFPE samples at the end of 2008 and has likely more than doubled by now, on top of serving many other satellite locations. Given that an average whole slide image at 80,000 x 60,000 pixels can take up 15 gB in full RGB, an entire databank of this size can span hundreds of petabytes of storage space, for one tissue bank in one city. Even storing these images in a JPEG compression standard would only yield ~10x size compression (Niazi et a., 2019). This is a problem for the transfer of data as well, as moving petabytes of data to and from more remote hospitals can take up a significant amount of time. Loading such data into memory for model training would also be difficult and would hamper researchers' ability to develop classification and regression models developed for these large, highly variable tissue datasets.

One solution proposed is the use of deep learning models for both the compression and search of these large tissue datasets. The development of such a model is certainly a difficult task as compression loss negatively effects model performance and even the highest performing compression model for the ImageNet database cannot even halve the file size of the JPEG2000 standard (Zhang et al., 2021). Reducing compression loss may be more realistic in WSI as, in 2018, a paper that suggested using importance maps in the deepest layers of the res-net was able to achieve a Peak Signal to Noise ratio (PSNR) of 29.2 and a Multi Scale Structural Similarity (MS-SSIM) of 0.92 which was more than JPEG and comparable to BPG (Alexandre et al., 2018). However, they did this while retaining more important features of the image at a lower Bits Per Pixel (bpp) rate when compared to JPEG and BPG, which means a matched fidelity with lower compression then the best non-deep learning-based compression algorithm. Most recently, learning models have scored MS-SSIMs above 0.97 using a combination of autoencoders, GANs, and FCNNs (Li et al., 2022, Ma et al., 2021, Zhao et al., 2021).

While CNN encoders have been used successfully in subsequent classification tasks of encoded whole slide images, lower bit representations of images maintain very high reconstructed loss (Aswolinskiy et al., 2021). Alternatively, autoencoders have long been a primary method of lossy image compression through neural networks because of their adaptive encoder-decoder structure which allows for high PSNR in reconstructed images at lower bit depths (Tan et al., 2009; Theis et al., 2017; Cheng et al., 2018). Studies have already shown that autoencoders retain high tissue fidelity at very high compression ratios (> 1:64) and high predictive power in classification tasks (Hecht et al., 2020; Lomacenkova and Arandjelović, 2021).

The use of variational autoencoders has recently become appealing for its use in image data compression due to its potential for Bayesian inference reduction of variation between encoded pixels to reduce loss during compression (Ballé et al., 2018; Zhou et al., 2018). In a comparative study of autoencoders to variational autoencoders (VAE), Sun et al. (2021) demonstrated that Variational Autoencoders (VAE) could possibly achieve slightly higher PSNRs at the cost of increased size and longer computational time. We saw this directly in ours and others application of VAEs to the compression of cancer H&E pathology images (Fig 3) (Nasr et al., 2023 and Way and Green, 2018). An in-depth comparison of multiple compressive models found that models trained on VAE encoded pixel space draw on information from the entire tissue image as opposed to parts of the image or individual cells for BiGAN and contrastive models (Tellez et al., 2019). Perfecting a compressed space encoding for cancer imaging data will allow researchers and clinicians to easily transfer and store large gigabyte sized microscopy images and even search a compressed space for similar patients (Chen et al., 2022 and Nasr et al., 2023).

## Clinical Application and Barriers of Adoption

Surveys show that the majority of clinicians believe that AI will lead to improved efficiencies and diagnostic accuracy and yet, the adoption and translation of these tools into clinical practice has been slow (Sarwar et al., 2019). Despite hundreds of FDA-approved computer-vision technologies being available on the market, the majority of practicing radiologists have yet to integrate these tools into their workflow. A recent survey of US radiologists revealed that only 30% are using AI-based tools in their workflow (Allen et al., 2021). This delayed adoption might be in part due to a combination of ethical, technical, and financial barriers to the application of the technology.

Of concern is how AI models could open medical practitioners to liability from incorrect diagnosis or poor pharmaceutical and surgical treatment. Of the 46 studies mentioned in this study (Table 1), 10 report an improved diagnostic accuracy over a human pathologist. While this would indicate that some models produced can reduce Type I and Type II errors, there is always going to be an amount of risk, especially when generalized to new data types or tissues that the model wasn't originally intended to handle. Most studies of diagnostic tool use find that the high reliance of low-value tools open a practice to liability from unnecessary testing or harm to the patient (Müskens et al., 2022). More effective may be models that are developed to assist pathologists to increase their performance or efficiency rather than those developed to replace pathologists entirely. For example, in a study where a pretrained dense net was given to pathologists to help them classify liver cancer hepatocellular carcinoma, the model did not improve overall accuracy but did increase the accuracy of a subset of more experienced pathologists (Kiani et al., 2020). The authors warn that poorly trained models can have a negative impact on diagnostic accuracy and should be taken into consideration when developing commercial solutions. Most of the current FDA-approved computer vision tools fit into this category of CAD and are likely to be used by pathologists in the lab whose expertise is unlikely to be completely replaced any time soon.

While many have rightly identified the ethical and technological obstacles involved, less attention has been paid to the lack of economic incentives for both clinicians and health systems to adopt AI-based tools within the fee-for-system environment of American medicine (Esteva et al., 2021; Cui et al., 2021; Gerke et al., 2020; and Singh et al., 2020). In fact, per a recent survey by KPMG of healthcare CEOs, despite their immense interest in AI, many felt that the high upfront costs of implementing AI were a major barrier, suggesting the need for improved return on investment (KPMG, 2020). While improving diagnostic accuracy may at first seem an obvious value proposition to health systems and providers, this is not necessarily true in a fee-for-service system where virtually all activities related to testing and treatment, including those associated with errors, are still reimbursable.

Others have suggested that AI will bring about improved efficiencies that will generate returns to health systems; however, by replacing costly human work, these technologies may actually reduce reimbursement and thus healthcare salaries over the long run. As a hypothetical example, imagine a product that automates the diagnosis of cancer, only requiring pathologists to manually review outlier cases. Once widely adopted, payers will rightfully argue that the reimbursement of AI-based cancer diagnoses should be lower than equivalent human efforts due to improved cost-efficiencies, as already proposed by some think tanks (Abràmoff et al., 2022). Under these circumstances, pathologists and the health systems that employ them may see a reduction in income on a per unit basis unless AI can meaningfully increase provider productivity to make up for any lost revenues. Should this happen, clinicians and health systems will now be responsible for many more cases at any given time, which paradoxically increases each parties' medico-legal duties and responsibilities. As a real-life example, in 2021 the Centers for Medicare & Medicaid Services (CMS) began reimbursing for AI-based digital retina exams at $45-65 per exam, which is approximately 50% less than human equivalent work (CMS, 2021).

Despite misaligned economic incentives and other obstacles not addressed here that might hinder the rate of adoption, there are more reasons to be optimistic about AI in cancer imaging over the long run. As many have already pointed out, AI has the potential to make a meaningful dent in some of the thorniest problems in medicine such as reducing waste and expanding access to high quality health services.

## Conclusion

The advent of AI in cancer tissue imaging signifies a groundbreaking shift in the realm of medical diagnostics and research. This study elucidated five core tasks central to this evolution, namely regression, classification, segmentation, generation, and compression tasks, each serving a unique function in the advancement of cancer detection and treatment. This paper elucidated that these AI algorithms not only exhibit immense potential in boosting the efficacy of cancer pathology diagnostics but also face a set of technical and ethical challenges that need to be surmounted for seamless integration into existing systems. The rapid advancement of AI promises a revolution in tissue pathology, potentially solving issues of low consensus and subjective human determination of disease. Despite the initial success, the present state of AI application in tissue pathology is in its nascent stage (Hajighasemi et. al., 2023; Saurav et. al., 2023; Nasr et. al., 2023; Alsup et. al., 2023), indicating the need for continued research and development. Future experiments in this domain, building on the insights offered in this review, could significantly augment the speed, accuracy, and reliability of cancer diagnosis and treatment, ultimately translating into improved patient outcomes. As we delve deeper into the uncharted territories of AI's capabilities, it becomes increasingly clear that we stand on the precipice of a significant turning point in medical science, one that could redefine cancer diagnostics and care.

## Data and Software Availability

No data was generated for this review.

## Author Contributions

M.R and J.L were responsible for conceptualization for this study. A.M, M.S.N, J.P.V., A.M.A, B.R., H.S, K.F., P.B., P.K., and S.J.R contributed to resource development. M.R., A.M., M.S.N. and J.P.V contributed to visualization. M.R. and J.L. wrote and edited the manuscript.

## Competing Interests

No competing interests are disclosed by the authors.

## Grant Information

This work was supported by a University of Texas System Rising STARs Award (J.M.L) and the CPRIT First Time Faculty Award (J.M.L).

Zhu, Xinliang, Jiawen Yao, and Junzhou Huang. "Deep convolutional neural network for survival analysis with pathological images." 2016 IEEE International Conference on Bioinformatics and Biomedicine (BIBM). IEEE, 2016.

**Figures and Tables**

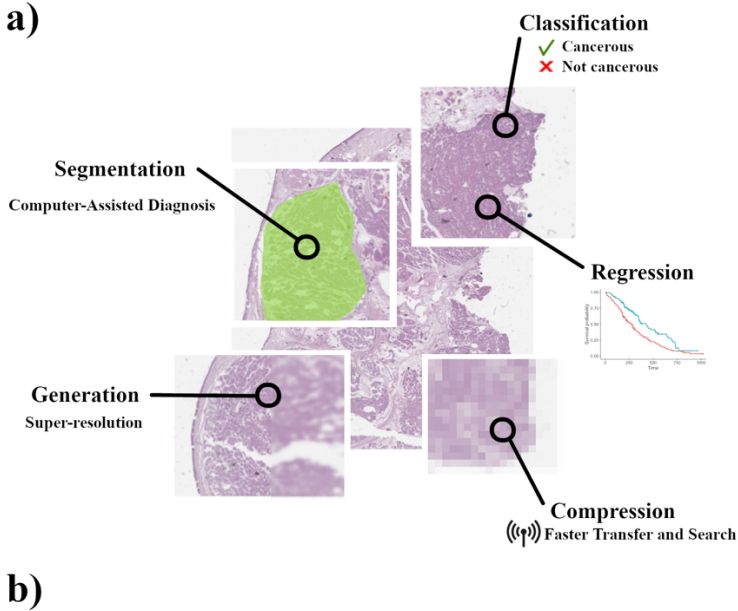

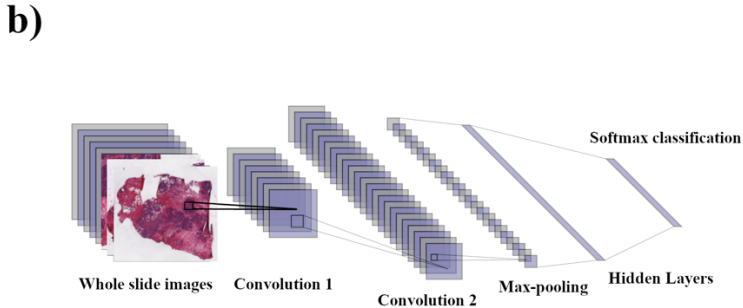

Figure 1. Overview of AI application in histopathology imaging. a) Example of the application of models for the 5 tasks in pathological slide analysis and application. b) Schematic overview of FCN when applied to WSI.

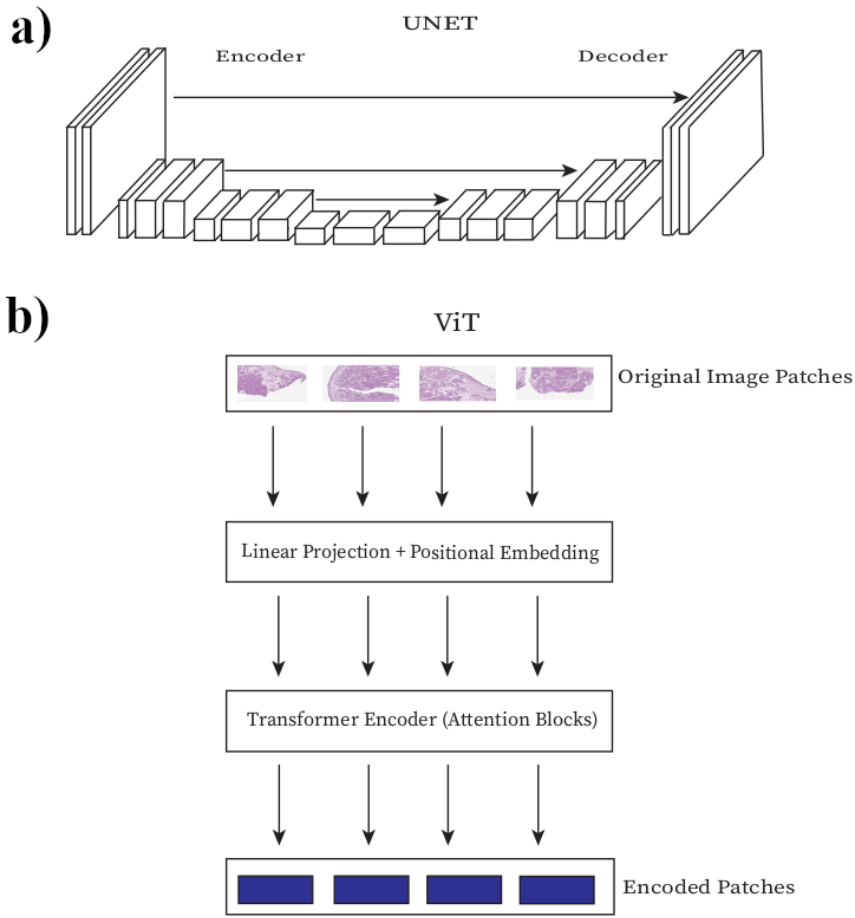

Figure 2. Common deep learning architectures for processing WSI input. a) U-Net architecture b) Visual Transformer architecture. Boxes represent layers in the network connected to adjacent layers.

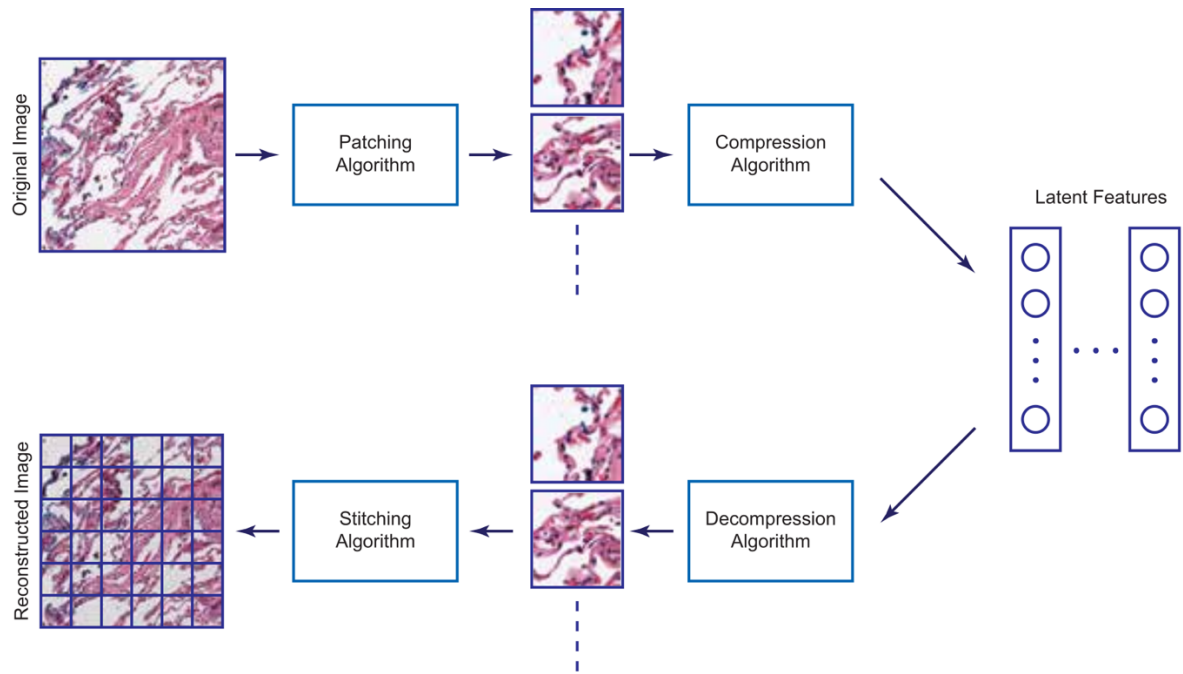

Figure 3. Representation of compression model architecture. Latent features are held in compressed space dependent upon bit depth.

Table 1. Summary of models cited in this review. Greatest accuracy claimed was reported but is not constant for dataset across studies. ROC/AUC included when reported.

| Model | Source | Accuracy | ROC/AUC | Tissue / Cancer | Architecture | Data Type |
|---|---|---|---|---|---|---|
| *Classification* | | | | | | |
| CNN | Nir et al. 2018 | 90.50% | 0.85 | Prostate Cancer | U-net | H&E |
| BNN | Syrykh et al. 2020 | 91% | 0.99 | Lymphoma | Bayesian Neural Network | H&E |
| CNN | Aichinger et al., 2017 | 93.6%% | -- | Liver Cancer | AlexNet | H&E |
| CNN | Mishra et al. 2018 | 92% | -- | Osteosarcoma | AlexNet | H&E |
| CNN | Sharma et al., 2017 | 69.90% | -- | Gastric Carcinoma | AlexNet | H&E |
| Unsupervised clustering, SVM | Di Cataldo et al., 2008 | 90% | -- | Lung Cancer | K-Means Clustering | IHC |
| RF/CNN | Cruz-Roa et al. 2014 | 84.23% | -- | Invasive Ductal Carcinoma (IDC) | CNN | H&E |
| CNN | Coudray, et al., 2018 | 97% | 0.97 | Lung Cancer | Inception V3 | H&E |
| CNN, RNN | Campanella et al., 2019 | 98% | 0.99 | Prostate, skin and axillary lymph nodes | ResNet34, AlexNet, VGG11 | H&E |
| CNN | Kiani et al., 2020 | 88.50% | -- | Liver Cancer | DenseNet | H&E |
| CLAM | Lu et al. 2021 | 90.20% | 0.951 | Kidney, Lung, Lymph Node, Breast Cancer | Transfer learning and CNN, Attention-based learning | H&E |
| CNN, MIL | Wetstein et al., 2022 | 80% | -- | Breast Cancer | ResNet-34 pretrained on ImageNet | H&E |
| CNN, Deep Learning System (DLS) | Nagpal et al., 2019 | 70% | 0.96 | Prostate Cancer | Inception V3, K-NN | H&E |

| Model | Reference | Accuracy | AUC | Cancer Type | Architecture | Data Type |
|---|---|---|---|---|---|---|
| LSTM-RNN | Poojitha et al., 2019 | 98.00% | 1 | Prostate Cancer | VGG-16, DCGAN, LSTM, RNN | H&E |
| DeepGrade | Wang et al., 2022 | -- | 0.95 | Breast Cancer | Deep CNN, Inception V3 | H&E |
| CNN, RNN | Cheng et al., 2021 | 98.80% | 0.983 | Cervical Cancer | ResNet50, RNN | H&E |
| AI-based model | Harmon et al., 2020 | 84.60% | 0.866 | Lymph Nodes, Bladder Cancer | ResNet-101, CNN, DBSCAN | H&E |
| CNN | Kwak et al., 2017 | -- | 0.974 | Prostate Cancer | MatConvNet | TMAs H&E |
| CNN | Hou et al., 2016 | 83% | 0.908 | Lung Cancer and Glioma | ReLU + LRN | H&E |

*Regression*

| Model | Reference | Accuracy | AUC | Cancer Type | Architecture | Data Type |
|---|---|---|---|---|---|---|
| CNN | Wang et al., 2018 | 89.80% | -- | Lung Cancer | Inception V3 | H&E |
| DeepConvSurv | Zhu et al., 2016 | -- | -- | Lung Cancer | Deep Convolutional Neural Network | CT |
| CNN, Deep Learning System (DLS) | Wulczyn et al., 2020 | 95% | -- | 33 different cancer types | MobileNet CNN Architecture | H&E |
| DeepOmix | Zhao et al., 2021 | -- | -- | 33 different cancer types | Deep Learning Framework | Multi-omics data |
| MultiSurv | Vale-Silva et al., 2021 | 82.20% | -- | 33 different cancer types | ResNeXt-50, CNN, ImageNet | Multi-omics data |
| CNN, Deep Learning System (DLS) | Wulczyn et al., 2021 | 87.0-95.5% | 0.985 | Colorectal Cancer | MobileNet CNN Architecture, Inception V3 | H&E |

Segmentation

| Model | Reference | Accuracy | AUC | Cancer Type | Architecture | Data Type |
|---|---|---|---|---|---|---|
| UBCNN | Lu, et al., 2021 | 83.10% | -- | Breast Cancer | UNet, VNet | H&E |
| FCN | Yi et al., 2018 | 95.20% | -- | Lung Cancer | Caffe, VGG-16 | H&E |

| Method | Reference | Accuracy | Dice | Cancer Type | Models | Image Type |
|---|---|---|---|---|---|---|
| CNN | Vu et al., 2019 | 78% | -- | Glioblastoma multiforme (GBM), Lower grade glioma (LGG), Head and neck squamous cell carcinoma (HNSCC), and Non-small cell lung cancer (NSCLC) | DRAN, U-Net, FCN, ResNet50 | H&E Tiles |
| Encoder-Decoder Network | Wazir et al., 2022 | 98.07% | 0.693 | Kidney, Lung, Colon, Breast, Bladder, Prostate, Brain, Liver, Stomach | DCNN, U-Net, FCN, Masked-RCNN | H&E |
| DuAT | Tang et al., 2022 | 94.80% | 0.924 | Colon, Skin Cancer | Vision Transformers, PVT, GLSA, SBA | Colonoscopy images, Dermoscopic images, and microscopy images |
| CellSeg | Lee et al., 2022 | 71.20% | 0.7121 | Colorectal Cancer | Mask R-CNN | CODEX images |
| encoder-decoder based FCN | Khened et al., 2021 | 75% | 0.909 | Breast, Colon, Liver Cancer | DenseNet-121, Inception-ResNet-V2, DeeplabV3Plus | H&E |

*Generation*

| Method | Reference | Accuracy | Dice | Cancer Type | Models | Image Type |
|---|---|---|---|---|---|---|
| GAN | Baur et al., 2018 | 71.20% | -- | Melanoma | DCGAN, LAPGAN | Dermoscopic images |
| SSIM guided cGAN | Saurav et al., 2022 | -- | -- | Lung cancer | U-Net, GAN | CODEX images |
| StylePix2Pix | Toda et al., 2022 | 48.70% | -- | Lung cancer | pix2pix, U-Net, GAN, StyleGAN | CT |
| Conditional infilling GAN (ciGAN) | Wu et al., 2018 | -- | 0.896 | Breast Cancer | ResNet-50, GAN, VGG-19, ImageNet | Mammography |
| CGAN | Wong et al., 2020 | 77.35% | -- | Breast Cancer | GAN, CNN | H&E |
| GAN | Ahmad et al., 2022 | -- | 0.95 | Retinal, Skin, Brain, Cardiac Cancer | ResNet34, SRCNN, VGG19 | Retinal images, Dermoscopic Images, Brain MRI, 2D |

| Method | Reference | Accuracy | Score | Cancer Type | Architecture | Stain |
|---|---|---|---|---|---|---|
| | | | | | | Cardiac Ultrasound images |
| ESRGAN | Nneji et al., 2021 | 92% | 0.9887 | Breast Cancer | GAN, Siamese CNN | H&E |
| WA-SRGAN | Shahidi, 2021 | 99.11% | 0.962 | Breast Cancer, Lymph Nodes | ResNeXt-101, SRGAN, VGG19 | H&E |

*Compression*

| Method | Reference | Accuracy | Score | Cancer Type | Architecture | Stain |
|---|---|---|---|---|---|---|
| VAE | Nasr et al., 2023 | -- | -- | Brain, Breast, Bronchus and Lung, and Colon | VAE | H&E |
| Neural Image Compression (NIC) | Tellez et al., 2019 | 89.90% | 0.725 | Lymph Node, Breast, Rectum | VAE, BiGAN, CNN | H&E |
| Tybalt | Way et al., 2018 | -- | -- | 33 different cancer types | VAE, CNN, GAN | RNA-Seq |
| Neural Image Compression (NIC) | Aswolinskiy et al., 2021 | -- | 0.98 | Lung | DenseNet, CLAM, ImageNet, Encoder | H&E |
| Disentangled Autoencoder | Hecht et al., 2020 | -- | -- | Colon | Autoencoder, GAN, ResNet V2 | H&E and IHC |
| SISH | Chengkuan et al., 2022 | -- | 0.958 | Lung, Breast, Colon and Prostate | VQ-VAE | H&E Tiles |